\documentclass[aps,prl,letterpaper,11pt,twoside,tightenlines,nofootinbib,showpacs,preprint,twocolumn]{revtex4}
\usepackage{graphicx}
\usepackage[sort&compress]{natbib}
\usepackage{latexsym}
\usepackage{epsfig}
\usepackage{amsmath}

\def\be{\begin{equation}}
\def\ee{\end{equation}}

\newcommand{\ket}[1]{\left|{#1}\right\rangle}

\newcommand{\braket}[2]{\left\langle{#1}| #2 \right\rangle}
\newcommand{\ketbra}[2]{ | #1\rangle \langle #2 |}
\newcommand{\mean}[1]{\left\langle #1 \right\rangle}
\newcommand{\1}{\mathbf{1}}

\begin{document}

\title{Phase Space Non-commutativity and its Stability}
\author{Paolo Castorina$^{1,2}$, Alfredo Guerrera$^{1}$ and Tomislav Prokopec$^{3}$}
\affiliation{
\mbox{${}^1$ Dipartimento di Fisica, Universit\`a di Catania, Via Santa Sofia 64,
I-95123 Catania, Italy.}\\
\mbox{${}^2$ INFN, Sezione di Catania, I-95123 Catania, Italy.}\\
\mbox{${}^3$Institute for Theoretical Physics, Spinoza Institute \& EMME $\Phi$,} \mbox{Faculty of Science, Utrecht University,}
\mbox{Postbus 80.195, 3508 TD Utrecht, The Netherlands.}
}

\date{\today}
\begin{abstract}

We consider a generalised non-commutative space-time in which non-commutativity is extended to
all phase space variables. If strong enough, non-commutativity can affect stability of the system. 
We perform stability analysis on a couple of simple examples and show that a system can be 
stabilised by introducing quartic interactions provided they satisfy phase-space copositivity.
In order to conduct perturbative analysis of these systems one can use either canonical methods 
or phase-space path integral methods which we present in some detail.   

\end{abstract}
 \pacs{04.20.Cv,11.10.Wx,11.30.Qc}
 \maketitle

\section{1. Introduction}

Recently the so-called higher derivative theories have been carefully studied due to their possible role in the renormalizability of non local theories~\cite{Julve:1978xn} such as String Theory~\cite{Eliezer:1989cr,Hata:1988wn,Hata:1989ie}, Loop Quantum Gravity~\cite{Liu:2017puc} and non-commutative field theories~\cite{Connes:1997cr,Seiberg:1999vs,Seiberg:2000gc}. 

As well known, higher derivative theories have serious constraints on their physical viability~\cite{Schmidt:1994iz} 
and the Ostrogradsky Instability theorem classifies all nondegenerate higher derivative theories as unstable. More precisely, the Ostrogradsky argument relies on having the highest momentum associated with the highest derivative in the theory and the  energy spectrum results unbounded from below. The physical system is therefore unstable and, after quantization, negative norm states (ghosts) usually appear~\cite{Deser:1974cz}
 and  the theory is not unitary~\cite{Stelle:1977ry}. For an interesting approach see~\cite{Simon:1990ic}.  

For that reason such theories have often been considered as effective field theories and, in general,  
one  assumes that the parameters multipling the higher derivative terms  are small, justifying 
perturbative treatment of higher derivative operators~\cite{Muthukumar:2002}.

Another way of tackling higher derivative theories is to introduce 
a Lorentz covariant cutoff operator~\cite{Kimberly:2003hp}.
For certain class of such theories, the propagator poles in momentum space shift but the number of poles does not change. 
Barnaby and Karman~\cite{Barnaby:2007ve} claim to have proven a theorem that states that theories 
in which the number of poles is preserved do not contain Ostrogradsky's  instabilities. 
While this statement is intriguing, in the absence of complete analysis of the constraint structure of the theory {\it \`a la} Dirac,
it is fair to say that a general proof is still lacking.

In this paper we consider a more traditional approach to non-commutativity, which introduces Lorentz breaking non-commutative parameters in phase-space. 
An important advantage of this class of models is the disappearance of the instabilities present in Lorentz covariant approaches. 
Namely, in the limit of strong non-commutativity, tachyonic modes appear, thus destabilising the theory. 
However, these kind of tachyonic instabilities are easy to deal with.  
To stabilize the theory it suffices 
to introduce interaction terms which have the property of copositivity. 

 Motivated by gravity coupled to (scalar) matter, which contains momenta in 
vertices~\cite{Maldacena:2002vr, Prokopec:2012ug},
our interaction terms are bi-quadratic in both position and momentum variables.
A natural framework to perform perturbative calculations in such theories is 
the {\it phase space path-integral formalism} (PSP), which we develop here for this purpose. 
We apply PSP to define general perturbation expansion, and 
show how to apply it in a simple one-loop case. We also check that the PSP and canonical method give identical answers.

The paper is organized as follows. 
In section~2\ref{sec:freetheory} the phase space path integral approach is discussed. The non-commutative harmonic oscillator is recalled in  
section~3 and section~4 is devoted to a pertrurbative study of interactions in the non-commutative harmonic oscillator.
Section~5 contains our final comments.

\section{2. PSP technique}
\label{sec:freetheory}

In this section we recall some preliminary definitions, by the direct application to the simple quantum harmonic oscillator with Hamiltonian
\begin{equation}
H=\frac{p^2}{2m}+\frac{1}{2}m\omega^2 x^2,\label{0}
\end{equation}
and then we discuss the phase space path integral formalism.
\subsection{a. Preliminary definitions}

The Feynman propagator defined as
\begin{equation}\begin{split}
\mean{\mathrm{T}[x(t)x(t^\prime)]}=&\theta(t-t^\prime)\mean{x(t)x(t^\prime)}+\\&+\theta(t^\prime-t)\mean{x(t^\prime)x(t)}, \label{2}\end{split}
\end{equation}
obeys the simple differential equation
\begin{equation}
(-\partial^2_t-\omega^2)\mean{\mathrm{T}[x(t)x(t^\prime)]}=\frac{i\hbar}{m}\delta(t-t^\prime),\label{1}
\end{equation}
whose solution depends on the choice of the initial state.
The coefficients of the two $\theta$ functions in Eq.~(\ref{2}), called \emph{Wightman functions}, are complex conjugate, i.e.
\begin{equation}\begin{split}
&i\Delta^+(t;t^\prime)=\langle x(t)x(t^\prime)\rangle,\\& i\Delta^-(t;t^\prime)=\langle x(t^\prime)x(t)\rangle,\end{split}
\end{equation}
with 
\begin{equation}
i\Delta^+(t;t^\prime)=[i\Delta^-(t;t^\prime)]^*,
\end{equation}
and satisfy the homogeneous equations
\begin{equation}
(-\partial_t^2-\omega^2)i\Delta^\pm(t;t^\prime)=0.
\end{equation}
If the initial state is the vacuum state one has
\begin{equation}
i\Delta^+(t;t^\prime)=\frac{\hbar}{2 m\omega}e^{-i\omega (t-t^\prime)}
\end{equation}
and 
\begin{equation}
i\Delta^-(t;t^\prime)=\frac{\hbar}{2 m\omega}e^{i\omega (t-t^\prime)}.
\end{equation}

Let us study other T-ordered products that we shall use later.  In particular, $\mean{\mathrm{T}[p(t)p(t^\prime)]}$ satisfies the differential equation (derived in appendix A)
\begin{equation}
\!\left(\!-\frac{d^2}{dt^2}\!-\!\omega^2\right)\mean{\mathrm{T}[p(t)p(t^\prime)]}
     = i\hbar m\omega^2 \delta(t\!-\!t^\prime).\!
\label{eq:pp}
\end{equation}
To evaluate the other T-ordered product 
\begin{equation}
\mean{\mathrm{T}[p(t)x(t^\prime)]},\label{10a}
\end{equation}
one starts from the operator solution of the equation of motion associated with
the Hamiltonian~(\ref{0}), i.e.
\begin{equation}\begin{cases}
q(t)=q_0 \cos \omega t+ \frac{p_0}{m\omega} \sin \omega t\\
p(t)=p_0 \cos \omega t - m\omega q_0\sin \omega t
\end{cases},
\end{equation}
and under the conditions
\begin{equation}
\begin{cases}
\mean{q_0^2}=\frac{\mean{p_0^2}}{(m\omega)^2}\\
\mean{\{q_0,p_0\}}=0
\end{cases},
\end{equation} 
one obtains
\begin{equation}\label{pq}\begin{split}
\mean{\mathrm{T}[p(t)x(t^\prime)]}=&-\theta(t-t^\prime)\frac{i\hbar}{2}e^{-i\omega(t-t^\prime)}+\\&+\theta(t^\prime-t)\frac{i\hbar}{2}e^{i\omega(t-t^\prime)}.
\end{split}\end{equation}

These are the two point functions of the ground state
of the simple harmonic oscillator~(\ref{0}).
The two point functions of more general states such 
as general gaussian states are discussed e.g. in 
Ref.~\cite{Glavan:2018}.

\subsection{b. Phase space path integral formalism}
The formalism of phase space path integral requires a detailed discussion of  mixed representations. The partition function between an initial state $\ket{\psi_0}$ at $t_0$ and a final state $\ket{\psi}$ at $t_*$ is defined~\cite{Ryder:1985wq} as 
\begin{equation}\begin{split}
&Z[J_i]=\braket{\psi,t_*}{\psi_0,t_0}=\Bigg\langle\psi\Big|\mathrm{T}\Big[\exp\Big(-\frac{i}{\hbar}\times\\&\int^{t_*}_{t_0} dt [H(t)-J_q q(t)-J_p p(t)]\Big)\Big]\Big|\psi_0\Bigg\rangle,\label{Zpartition}
\end{split}\end{equation}
where $J_q$ and $J_p$ are different sources, coupled with position and momentum respectively.
By completeness of $q$ and $p$ representations, the second term can be written as
\begin{equation}\begin{split}
\int dq_* dq_0 dp_*& \braket{\psi}{q_*}\braket{q_*}{p_*}\times\\&F_J(p_*,t_*|q_0,t_0)\braket{q_0}{\psi_0},
\end{split}\end{equation}
where $q_*$ and $p_*$ are the position and momentum of the system at $t=t_*,$ $q_0=q(t_0)$ and
\begin{equation}\begin{split}
&F_J(p_*,t_*|q_0,t_0)=\Big\langle p_*\Big|\mathrm{T}\Big[\exp\Big(-\frac{i}{\hbar}\times\\&\int^{t_*}_{t_0} dt [H(t)-J_q q(t)-J_p p(t)]\Big)\Big]\Big|q_0\Big\rangle.
\end{split}\end{equation}
Let us now consider a generic Hamiltonian $H$ of the form\begin{equation}
H=H_0+H_{\mathrm{Int}},
\end{equation}
where the exact solutions for $H_0$ are known.   
By dividing the time interval $t_*-t_0>0$ in $N$ small intervals and evaluating the $N\rightarrow \infty$ limit, it turns out ( see the detail in appendix B) that

\begin{equation}\begin{split}
Z[J_i]=&\int dq_* dq_0dp_*\times\\&\psi^*(q_*)\psi_0(q_0) \frac{e^{i\frac{p_*q_*}{\hbar}}}{\sqrt{2\pi\hbar}}F_J(p_*,t_*|q_0,t_0),\label{pfunction}
\end{split}\end{equation}
which can be written as
\begin{equation}\begin{split}
Z[J_p,J_q]=\exp\Bigg(\frac{i}{\hbar}S_{\mathrm{int}}\Big[\frac{h}{i}\frac{\delta}{\delta J_q},&\frac{h}{i}\frac{\delta}{\delta J_p}\Big]\Bigg)\times\\&Z_0[J_p,J_q],\label{pert.exp}
\end{split}\end{equation}
where 
$$Z_0\propto \int \mathcal{D}q\mathcal{D}p e^{\frac{i}{\hbar}\int_{t^0}^{t^*}dt\{p\dot{q} - H_0(p,q)+J_qq(t)+J_pp(t)\}},$$
and the momenta are explicitely taken into account.
In the previous expressions $\mathcal{D}q$ and $\mathcal{D}p$ indicate the usual functional integration measure:\begin{equation}
\mathcal{D}q\mathcal{D}p =\lim_{N\rightarrow \infty}\Pi_{l,m=1}^{N-1}dq_l\frac{dp_m}{\sqrt{2\pi\hbar}},\end{equation} where $q_l=q(t_l),$ $p_m=p(t_m)$ with $t_l$ and $t_m$ belonging to $]t_0,t_*[.$ 

The presence of the source  terms, $J_p$ and $J_q$ , permits to set up a perturbative expansion for interaction terms that are not diagonal in position operators and contain momentum dependence or for Hamiltonians with non-canonical kinetic terms.

In the following we will use $H_0$ in Eq.~(1), with the corresponding action:
\begin{equation}\label{eq:6}\begin{split}
S_0 =&\int_{t^0}^{t^*}dt\Big ( p\dot{q}-\frac{p^2}{2m}-\frac{m\omega^2}{2}q^2  \Big ). \end{split}
\end{equation} 
which can be written as
\begin{equation}\label{eq:7}\frac{1}{2}\int_{t^0}^{t^*}dt \int_{t^0}^{t^*}dt^\prime \Big [ Q_i(t) D_0^{ij}(t;t^\prime)Q_j(t^\prime)\Big ]\end{equation}
where $Q_1(t)=q(t),$ $Q_2(t)=p(t)$ and 
\begin{equation}\label{eq:8} \begin{split}&D_0^{ij}(t;t^\prime)=\\
&\left(
\begin{array}{cc}
-m\omega^2 & -\partial_t \\
\delta (t-t^*)- \delta(t-t^0) +\partial_t & -\frac{1}{m} \\
\end{array}
\right)\delta (t-t^\prime). \end{split}\end{equation}
The next step is to evaluate the inverse two point function after integrating out $p_*$ in Eq.~(\ref{pfunction}) with $H=H_0$.
One gets
\begin{equation}\begin{split}\label{eq:9}Z_0&[J_p,J_q]=\int dq^*\int dq^0\hspace{.5 ex}\rho (q^0,q^*)\times\\&\int _{q(t^0)=q^0}^{q(t^*)=q^*}\mathcal{D}q\mathcal{D}p\hspace{.5 ex}\mathrm{exp}\bigg \{\frac{i}{\hbar} \int_{t^0}^{t^*} dt [p\dot{q}+\\&\qquad\qquad\quad - H_0(p,q) + J_pp+J_qq]\bigg \},\end{split}\end{equation}
with \begin{equation}\label{eq:10}\begin{split}
\braket{q_0}{\psi_0}&\braket{\psi}{q_*}=\rho (q^0,q^*)=\\& \sqrt{\frac{m\omega}{2\hbar}}\exp \Big \{-\frac{m\omega}{2\hbar}[(q^0)^2+(q^*)^2]\Big \}.
\end{split}\end{equation}
The direct substitution of Eq.~(\ref{eq:10}) in Eq.~(\ref{eq:9}) gives (b.c.=\emph{boundary conditions})
\begin{equation}\begin{split}\label{eq:11}Z_0&[J_p,J_q]=\int  dq^*\!\int dq^0\!\int _{\mathrm{b.c.}}\hspace{-0.5 cm}\mathcal{D}q\mathcal{D}p\hspace{.5 ex}\mathrm{exp}\bigg\{\frac{i}{\hbar} S_0 + \\&\quad+\int dt \Big (\frac{i}{\hbar}\frac{im\omega}{2}(\delta (t-t^0)+\delta (t-t^*) )q^2+\\&\hspace{4.5 cm}+ J_pp+J_qq \Big )\bigg \},\end{split}\end{equation} and therefore the inverse two point function becomes
\begin{equation}\begin{split}\label{eq:12}
&\tilde{D}_0^{ij}(t;t^\prime)=\delta (t-t^\prime)\times\\
&\left(
\begin{array}{cc}
-m\omega^2 +im\omega(\delta (t-t^0) +\delta(t-t^*)) & -\partial_t \\
\delta (t-t^*)- \delta(t-t^0) +\partial_t & -\frac{1}{m} \\
\end{array}
\right).\end{split} \end{equation}
The $H_0$ part of the generating function can be now written as \begin{equation}\begin{split}\label{eq:13} &Z_0[J_i]=\int\mathcal{D}Q_i\hspace{.5 ex}\mathrm{exp}\frac{i}{\hbar}\int_{t^0}^{t^*}dt \int_{t^0}^{t^*}dt^\prime\times \\&\Big [\frac{1}{2} Q_i(t) \tilde{D}_0^{ij}(t;t^\prime)Q_j(t^\prime) +J^i(t)\delta(t-t^\prime)Q_i(t^\prime)\Big ] \end{split}\end{equation}
where $J^1(t)=J_q$ and $J^2(t)=J_p$.
The initial conditions on the state are now self contained, i.e.
$$\int dq^*\int dq^0 \int_{q(t^0)=q^0}^{q(t^*)=q_*} \mathcal{D}q\dots=\int \mathcal{D}q\dots$$
where $t_0<t_*.$

Shifting the $Q_i$'s via $Q_i(t)\rightarrow \tilde{Q}_i(t)=Q_i(t)+\frac{1}{\hbar}\int d\tilde{t}i\Delta_{ij}(t;\tilde{t})J^j(\tilde{t}),$
where $i\Delta_{ij}(t;t^\prime)$ satisfies the condition
\begin{equation}\label{eq:14} \int d\tilde{t}\tilde{D}_0^{ij}(t;\tilde{t})i\Delta_{jk}(\tilde{t};t^\prime)=i\hbar\delta^i_k\delta(t-t^\prime),\end{equation}
one obtains  (neglecting the $\sim$ for the integration variables) 
\begin{equation}\begin{split}\label{eq:15}
&Z_0[J_i]=\int \mathcal{D} Q_i\times\\& \exp \Big ( \frac{i}{\hbar} \int_{t^0}^{t^*}dt\int_{t^0}^{t^*}dt^\prime \frac{1}{2} Q_i(t) \tilde D_0^{ij}(t;t^\prime)Q_j(t^\prime) \Big )\times\\&\exp \Big ( -\frac{i}{2\hbar^2} \int_{t^0}^{t^*}dt\int_{t^0}^{t^*}dt^\prime J^i(t)i\Delta_{ij}J^j(t^\prime) \Big ).\end{split} \end{equation} 
Integrating out the $Q_i$'s we finally get \begin{equation}\label{eq:16}\begin{split}
&Z_0[J_i]=\tilde{Z}_0(t^0;t^*)\times\\&\hspace{-0.15 cm}\exp \Big (-\frac{1}{2\hbar^2} \int_{t^0}^{t^*}\hspace{-.1 cm}dt\int_{t^0}^{t^*}\hspace{-.1 cm}dt^\prime J^i(t)i\Delta_{ij}(t;t^\prime)J^j(t^\prime) \Big ), \end{split}\end{equation} where $\tilde{Z}_0(t^0;t^*)$ is a $J_i$-independent constant.

The crucial point is to solve Eq.~(\ref{eq:14}) for $i\Delta_{ij}(t;t^\prime)$, once $\tilde{D}_0^{ij}$ is known, in order to obtain the generating functional of the $H_0$ part, $Z_0[J_i].$ 
From now on we shall refer to the $H_0$ two point function  $\tilde{D}_0^{ij}$ as $D^{ij}$ and we focus on the problem of the inversion of Eq.~(\ref{eq:14}).
The solution (see appendix C) is 
\begin{equation}
i\Delta_{ij}(t;t^\prime)=\mean{T[Q_i(t)Q_j(t^\prime)]}\label{PSP},
\end{equation}
where all the phase space two-point functions are considered.
Although we worked out the computation for a specific system, this method should hold for generic Hamiltonians. Indeed, the phase space path-integral approach  keeps track of the degrees of freedom of the full symplectic manifold where the evolution takes place and it is a useful starting point for  perturbative calculations,  in particular for systems 
with derivative interactions.

\section{3. Non-Commutative oscillator \protect \\ in two dimensions}

In this section  section we introduce the non-commutative oscillator in two 
spatial dimensions (2D). The total Hamiltonian can be conveniently split into 
a free part $H_0$ and an interacting part $H_{\rm int}$. 

 The free part of the Hamiltonian is given by, 
\begin{equation}
 H_0 = \frac{ p^{\prime 2}_x}{2m}+\frac{ p^{\prime 2}_y}{2m}
+\frac{1}{2}m\omega^2( x^{\prime 2}+ y^{\prime 2})
,
\label{H 0}
\end{equation}
and the interaction Hamiltonian is, 
\begin{eqnarray}
  H_{\rm int} &=& \frac{\lambda_x}4{x'}^4
    + \frac{\lambda_y}4 {y'}^4
 + \frac{\lambda_{p_x}}4 {p'}_x^4
    + \frac{\lambda_{p_y}}4 {p'}_y^4
\;\;
\label{H int}\\
 &+& \frac{\lambda_{xy}}4 {x'}^2 {y'}^2
    \!+\! \frac{\lambda_{xp_x}}4 {x'}^2 {p'}_x^2
 \!+\! \frac{\lambda_{xp_y}}4 {x'}^2 {p'}_y^2
\nonumber\\
 &+&\! \frac{\lambda_{yp_x}}4 {y'}^2 {p'}_x^2
    \!+\!  \frac{\lambda_{yp_y}}4 {y'}^2 {p'}_y^2
 \!+\!  \frac{\lambda_{p_xp_y}}4 {p'}_x^2 {p'}_y^2
.\;\;\;
\nonumber
\end{eqnarray}
The Hamiltonian describes a stable theory if it is copositive with respect to 
all directions in phase space, $({x'}^2,{y'}^2,{p'}_x^2,{p'}_y^2)$ for large
values of the coordinates~\cite{Chataignier:2018kay,Chataignier:2018aud}, and we assume
that to be the case 
in this work.

 Canonical quantization posits that conjugate variables do not commute, 
\begin{equation}
[ x', p'_x]=i\hbar
\,,\quad 
[y', p'_y]=i\hbar
,
\label{canonical quantization}
\end{equation}
and all other quantities commute. 
A generalisation to non-commutative geometries has been studied a lot, as it is well motivated by the physics at large energies. Arguably the simplest model is 
the one in which spatial coordinates do not commuate, 
$[ x', y']=i\theta_{xy}\equiv i\theta$. 
Here we consider a phase space generalization of this, in which all coordinates 
on phase space $q'_i$ do not commute, including the momenta, 
\begin{equation}
  [ q'_i,q'_j]=i\theta_{ij}
\,,
\label{general noncommutativity}
\end{equation}
 where $q'_i \equiv \{ x', y', p'_x,p'_y\}$ and $\theta_{ij}$ 
 is an antisymmetric hermitean matrix, but otherwise not specified. 
Notable special cases are canonical quantisation, in which 
$\theta_{xp_x}=\theta_{yp_y}=\hbar$ and all other elements are zero,
and non-commutativity in the $xy$ plane in which, in addition, 
$\theta_{xy}\equiv \theta\neq 0$. In the general case there are 3 more 
  non-vanishing elements of $\theta_{ij}$: $\theta_{xp_y}$, $\theta_{yp_x}$
and $\theta_{p_xp_y}$. 

\subsection{3.a Non-commutativity in space} 

 Let us now consider a simple model in which the coordinates  satisfy 
the commutation relations
\begin{equation}
[x^\prime, y^\prime]=i\theta_{12},
\end{equation}
with
\begin{equation}
\theta_{12}=-\theta_{21}=\theta,
\end{equation}
and all the other commutators are the usual commutative ones (the indices 1 and 2 correspond to  $x$ and $y$ directions).
In terms of the usual commuting set of phase space variables, $\{x,  y, p_x, p_y\}$, 
 the non-commuting coordinates are given by  
(see Ref.~\cite{Carroll:2001ws})
\begin{equation}
\begin{cases}
 x^\prime= x-\frac{\theta}{2\hbar} p_y\\
 p^\prime_x=p_x\\
 y^\prime= y+\frac{\theta}{2\hbar} p_x\\
p^\prime_y=p_y,
\end{cases}\label{cj}
\end{equation}
and $H_0$ in terms of commutative variables turns out to be
\begin{equation}
\begin{split}
H_0=&\left(\frac{1}{2m}+\frac{m\omega^2 \theta^2}{8\hbar^2}\right)( p^2_x+p^2_y)\\
+&\frac{1}{2}m\omega^2( x^2+ y^2)-\frac{m\omega^2\theta}{2\hbar}(x p_y- y p_x),
\end{split}\
\end{equation}
Following Ref.~\cite{Muthukumar:2002} we define $M$ and $\Omega$ as
\begin{equation}
\frac{1}{2M}=\frac{1}{2m}+\frac{m\omega^2 \theta^2}{8\hbar^2}, \qquad M\Omega^2=m\omega^2,\label{space noncommutativity}
\end{equation}
so that $H_0$ takes the more familiar form:
\begin{equation}
\begin{split}
 H_0=&\frac{1}{2M}(p^2_x+p^2_y)+\frac{1}{2}M\Omega^2(x^2+y^2)
\\&-\frac{M\Omega^2\theta}{2\hbar}( x p_y- y p_x).
\label{hamiltonian}
\end{split}
\end{equation}
In order to gain understanding on the question of stability of this Hamiltonian, it is useful to rewrite~(\ref{hamiltonian}) as: 
\begin{equation}
\begin{split}
\label{hamiltonian:II}
H_0=&\frac{1}{2M}\left(p_x+\frac{M^2\Omega^2\theta}{2\hbar}y\right)^2
\\&        + \frac{1}{2M}\left( p_y-\frac{M^2\Omega^2\theta}{2\hbar}x\right)^2
\\& 
    +\frac{1}{2}M\Omega^2\left(1-\frac{M^2\Omega^2\theta^2}{4\hbar^2}\right)(x^2+y^2)
\,,
\end{split}
\end{equation}
from where we see that the Hamiltonian is stable (it is a sum of non-negative terms) if 
\begin{equation}
\theta\leq \theta_c = \frac{2\hbar}{M\Omega}
\,.
\label{stability}
\end{equation}
If $\theta =\theta_c$ it is marginally stable and if $\theta>\theta_c$ both $x$ and $y$ directions in $H_0$
 are unstable (tachyonic) in the sense that the oscillator frequencies 
squared are negative. This instability can be cured by adding to the free Hamiltonian
quartic interactions as in Eq.~(\ref{H int}) with $\lambda_x,\lambda_y>0$ and with other couplings in~(\ref{H int})
zero or positive.~\footnote{As already mentioned above, a weaker criterion of co-positivity sufficies to make the theory stable.} 

 The {\it classical} ground state of the sub-critical system with $\theta< \theta_c$ is then simply, $p_x=p_y=x=y=0$,
such that the ground state energy vanishes, $E_0^{\rm cl}=0$. 
In the critical case when $\theta=\theta_c$ the ground state of $H_0$ is given by,  
$p_x=-M\Omega(\theta/\theta_c) y_0$,
 $p_y=M\Omega(\theta/\theta_c) x_0$, with $x_0$ and $y_0$ arbitrary (in this case both $x$ and $y$ 
are flat directions such that it costs no energy to shift along  $x$ and $y$ directions). 
Also in this case the ground state energy is zero in classical theory, $E_0^{\rm cl}=0$.
Finally, when $\theta>\theta_c,$ both $x$ and $p$ condense via spontaneous symmetry breaking 
and the ground state is given by, 
\begin{eqnarray}
 x_0^2\! &=& \!\frac{M\Omega^2}{\lambda_x}\Big(\frac{\theta^2}{\theta_c^2}\!-\!1\Big)
,\;
 y_0^2 = \frac{M\Omega^2}{\lambda_y}\Big(\frac{\theta^2}{\theta_c^2}\!-\!1\Big)
, 
\nonumber\\
p_{x,0}&=&-M\Omega\frac{\theta}{\theta_c} y_0
,\;
p_{y,0}=M\Omega\frac{\theta}{\theta_c} x_0
\label{ground state coordinates}
\end{eqnarray}
and the correponding ground state energy is negative, 
\begin{equation}
 E_{0}^{\rm cl} = -\frac14M^2\Omega^4
         \left(\frac{1}{\lambda_x}+\frac{1}{\lambda_y}\right)
\,.
\label{gs energy}
\end{equation}
A similar result was obtained in the context of a 3D supersymmetric harmonic oscillator with spin non-commutativity, see Ref.\cite{spin.non.commutativity,Alavi:2005yr}.
We now pause to summarize what we have found so far. If non-commutativity in space is strong enough, it can induce 
condensation of coordinates that strongly resembles {\it spontaneous symmetry breaking,} 
which is one of the principal results of this work. 
Consequently, the ground state energy is negative and 
and the particle is moving in circles 
with a constant velocity $\|\vec v\|$,
which is in its ground state given by 
\begin{eqnarray}
 v_{x,0}=\frac{p_{x,0}}{M}=-\Omega^2\frac{\sqrt{M/\lambda_x}}{\sqrt{1-\theta_c^2/\theta^2}}
\nonumber\\
v_{y,0}=\frac{p_{y,0}}{M}=\Omega^2\frac{\sqrt{M/\lambda_y}}{\sqrt{1-\theta_c^2/\theta^2}}
\,,
\end{eqnarray}
{\it i.e.} as a consequence of the space non-commutativity, we will perceive the particle moving without any 
`apparent' reason. Furthermore, recalling the structure of the Hamitonian in an external Abelian gauge field 
$A_\mu \equiv (A_0,A_x,A_y,A_z)$, whose kinetic part reads,
\begin{equation}
 H_{0,\rm gauge} = \frac{(\vec p - (e/c)\vec A)^2}{2M}
\label{gauge Hamiltonian}
\end{equation}
we see that, as alredy shown in \cite{Jackiw:2002wd} and many other works, the noncommutativity in space can be modeled by a fictitious gauge field, 
\begin{equation}
A_x =-\frac{c}{e}\frac{M^2\Omega^2\theta}{2\hbar} y 
, \;
A_y = \frac{c}{e}\frac{M^2\Omega^2\theta}{2\hbar} x
\,,
\label{fictitious gauge}
\end{equation}
which yields constant magnetic field in the $z$ direction, 
\begin{equation}
B_z = -\frac{c}{e}\frac{M^2\Omega^2\theta}{\hbar}. 
\end{equation}
The corresponding equivalent Lorentz force ($\vec F_L=(e/c)\vec v\times \vec B$) is then,
\begin{eqnarray}
F_x&=&-\frac{p_y}{M}\frac{M^2\Omega^2\theta}{\hbar}=-2\Omega\frac{\theta}{\theta_c} p_y
\nonumber\\
F_y&=&\frac{ p_x}{M}\frac{M^2\Omega^2\theta}{\hbar}=2\Omega\frac{\theta}{\theta_c} p_x
,
\label{Lorentz Force}
\end{eqnarray}
which will cause particles to rotate in the $xy$ plane with
a cyclotron angular frequency, $\omega_c=(eB)/(Mc)$, given by, 
\begin{equation} 
\omega_{\rm c} 
= 2\Omega\frac{\theta}{\theta_c}
\,.
\label{cyclotron frequency}
\end{equation}

In the quantum mechanical case when $\theta=\theta_c$ and interactions are switched off, 
the problem can be reduced to the 
two oscillator problem in a constant magnetic field pointing orthogonally to the $xy$ plane. 
The energy of the ground and excited states are then famously given by the Landau levels, whose energy is quantized as, 
\begin{equation}
E_n = \hbar\omega_c\left(n+\frac12\right),\quad n= 0,1,2,\dots
\label{Landau levels}
\end{equation}
When $\theta<\theta_c$ the oscillators will in their ground state rest on average at zero, and one can show 
that the ground state energy is equal to, $E_0= \hbar\omega_c/2$. However, in their 
excited state they can exhibit cyclotron rotation in the $xy$ plane. For example, if both $x$ and $y$ 
oscillators are in a coherent state, then the center of a Gaussian wave function will exhibit cyclotron oscillations 
in the $xy$ plane.

 
\subsection{3.b Non-commutativity in phase space}

We can extend the non-commutative geometry to the phase space via the relations
\begin{equation}
[p^\prime_x,p^\prime_y]=i\theta_{p_xp_y},
\end{equation}
with
\begin{equation}
\theta_{p_xp_y}=-\theta_{p_yp_x}=\eta
,
\end{equation}
analogous as in eq.(\ref{general noncommutativity}).
%

%
All the other commutation relations are the same as in the case discussed previously. Here we are introducing phase space non-commutativity in a simple but unnatural way, note that the same result can be obtained via purely physical arguments\cite{Banerjee:2001zi}.


The homeomorphism
between regular and non commutative geometry can be expressed via the coordinate change
\begin{equation}
\begin{cases}
x^\prime=C\left(x-\frac{\bar\theta}{2\hbar}p_y\right)\\
p_x^\prime=C\left(p_x+\frac{\bar\eta}{2\hbar}y\right)\\
y^\prime=C\left(y+\frac{\bar\theta}{2\hbar}p_x\right)\\
p_y^\prime=C\left(p_y-\frac{\bar\eta}{2\hbar}x\right)
\end{cases}\label{general omeomorphism}
\end{equation}
with \begin{equation}
C=\frac{1}{\sqrt{1+\frac{\bar\theta\bar\eta}{4\hbar^2}}}
\;,
\bar\theta =\frac{\theta}{1+\frac{\theta\eta}{4\hbar^2}}
\;,
\bar\eta =\frac{\eta}{1+\frac{\theta\eta}{4\hbar^2}}
.
\end{equation}
If one substitutes eq.(\ref{general omeomorphism}) in eq.(\ref{H 0}) the free part of the Hamiltonian can be recast in the same form as in eq.(\ref{hamiltonian}) but with redefined parameters and variables, namely 
\begin{equation}
\begin{split} 
H_0=\frac{1}{2\tilde M}\left({p}_x^2+{p}_y^2\right)+\frac{1}{2}\tilde M\tilde {\Omega}^2({x}^2+{y}^2)+\\-\frac{\tilde M\tilde {\Omega}^2\tilde {\theta}}{2\hbar}\left({x}{p}_y-{y}{p}_x\right),
\end{split}
\label{H0 shifted theta eta}
\end{equation} 
where 
\begin{equation}
\begin{split}
& \tilde M = m\frac{1+\frac{\bar\theta\bar\eta}{4\hbar^2}}{1+(m\omega)^2\frac{\bar\theta^2}{4\hbar^2}}
\\
& \tilde M\tilde {\Omega}^2=\frac{m\omega^2}{1+\frac{\bar\theta\bar\eta}{4\hbar^2}}
\left(1+\frac{1}{(m\omega)^2}\frac{\bar\eta^2}{4\hbar^2}\right)
,\\
&\tilde {\theta}=\frac{\bar\theta+\frac{\bar\eta}{(m\omega)^2}}{1+\frac{1}{(m\omega)^2}\frac{\bar\eta^2}{4\hbar^2}}
.
\end{split}
\label{H0 shifted theta eta:2}
\end{equation}
The limit $\eta\rightarrow 0$ is well behaved and produces the results~(\ref{space noncommutativity}--\ref{hamiltonian}) obtained in the case of space non-commutativity.

Indeed, studying the stability of the system, one obtains eq.(\ref{hamiltonian:II}) in terms of these redefined parameters, and the critical value is now realized as
\begin{equation}
\tilde \theta =\tilde \theta_c=\frac{2\hbar}{\tilde M\tilde \Omega},
\end{equation} 
granting exactly the same considerations as before,
but with the suitably rescaled non-commutative parameter $\theta\rightarrow \tilde\theta$.

\subsection{3.c PSP Feynman rules for non-commutative oscillator}

In what follows we construct two-point functions for a simple Gaussian initial state in the presence of 
the free Hamiltonian~(\ref{H 0}). These correlators will be used for studying the interactions~(\ref{H int}) in perturbation theory.
Since our interactions are higher order in both position and momentum variables, 
the phase space formalism developed in section 2 is particularly suitable. 
For simplicity we consider here only the sub-critical case ($\theta<\theta_c$) 
and leave the interesting super-critical case for future work.

The Hamilton operator equations can be conveniently written in a matrix form as, 
\begin{equation}
\frac{d}{dt}\!\left(\!\begin{array}{c}
                              x \cr
                               y \cr
                               p_x \cr
                               p_y \cr
                         \end{array}
                 \!\right)
               \!=\! \left(\!\begin{array}{cccc}
                             0 & \frac{\Omega\theta}{\theta_c} & \frac{1}{M} & 0\cr
                             -\frac{\Omega\theta}{\theta_c} & 0 & 0 &  \frac{1}{M}\cr
                            -M\Omega^2 &0 &0 & \frac{\Omega\theta}{\theta_c}\cr
                             0 &  -M\Omega^2&-\frac{\Omega\theta}{\theta_c}& 0 \cr
                         \end{array}
                 \!\right)\cdot
                    \left(\!\begin{array}{c}
                               x \cr
                                y \cr
                               p_x \cr
                               p_y \cr
                         \end{array}
                 \!\right)
\!.
\label{EOM H0}
\end{equation}
Making an {\it Ansatz}, $ q_i\propto \exp(\lambda t)$, one obtains four eigenvalues for $\lambda$,
\begin{equation}
 \lambda_\pm^2=-\Omega^2\left(1\pm\frac{\theta^2}{\theta_c^2}\right)
.
 \label{eingevalues}
\end{equation}
This means that -- when $\theta<\theta_c$ -- the four fundamental solutions are oscillatory harmonic 
functions,~\footnote{When $\theta>\theta_c$, the fundamental solutions either exponentially 
grow or decay
in time with the rate given by $\kappa_\pm=\sqrt{-\lambda_\pm^2}
=\Omega\sqrt{(\theta/\theta_c)^2\pm 1}$. The exponentially growing behavior comes from   
the tachyonic mode mentioned above.}
\begin{equation}
 \cos(\Omega_+ t)
,\;\;
 \cos(\Omega_- t)
,\;\;
 \sin(\Omega_+ t)
,\;\;
 \sin(\Omega_- t)
,
\label{fundamental solutions}
\end{equation}
where 
\begin{equation}
 \Omega_\pm=\Omega\sqrt{1\pm\frac{\theta^2}{\theta_c^2}}
\,.
\label{fundamental frequencies}
\end{equation}
The general solution for $ x(t)$ is
\begin{equation}
\begin{split}
 x(t)=&\frac{ x_0}{2}(\cos\Omega_+t+\cos\Omega_-t)+\\
+&\frac{ y_0}{2}(\sin\Omega_+t-\sin\Omega_-t)+
\\ +&
\frac{ p_{x0}}{2M\Omega}(\sin\Omega_+t+\sin\Omega_-t)+
\\+&
\frac{ p_{y0}}{2M\Omega}(\cos\Omega_-t-\cos\Omega_+t)
\end{split}
\end{equation}
where $ x_0= x(0),$ $ y_0= y(0),$ $ p_{x0}= p_x(0),$ and $ p_{y0}= p_y(0).$ 
The solutions for the others dynamical variables can be found in Appendix D.

The Wightman functions evaluate to,
\begin{equation}\begin{cases}
i\Delta^+_{ij}(t;t^\prime)=\mean{Q_i(t)Q_j(t^\prime)}\\
i\Delta^-_{ij}(t;t^\prime)=\mean{Q_j(t^\prime)Q_i(t)}.
\end{cases}
\end{equation} 
Indeed, the diagonalization of $H_0$ \cite{Muthukumar:2002} allows to calculate the expectation values we need to determine $i\Delta_{ij}(t;t^\prime).$ The only non-zero expectation values turn out to be \begin{equation}\begin{split}
\mean{x^2_0}=\mean{y^2_0}=\frac{\hbar}{2M\Omega},\\ \mean{p_{x0}^2}=\mean{p_{y0}^2}=\frac{\hbar M\Omega}{2},\\ \mean{x_0p_{x0}}=\mean{y_0p_{y0}}=\frac{i\hbar}{2}.
\end{split}\end{equation}

Some comments on the two point functions are now in order. In the limit $\theta \rightarrow 0$, the two propagators
\begin{equation}\begin{split}
&i\Delta_{xx}^+(t;t^\prime)=\frac{\hbar}{4M\Omega}\left(e^{-i\Omega_+t}+e^{-i\Omega_-t}\right),\\
&i\Delta_{p_xp_x}^+(t;t^\prime)=\frac{\hbar M\Omega}{4}\left(e^{-i\Omega_+t}+e^{-i\Omega_-t}\right).
\end{split}
\end{equation}
reproduce the well known Wightman functions of the commutative case. Moreover the non-commutative parameter generates two distinct frequencies $\Omega_\pm$ and unusual correlations, as for  example 
\begin{equation}
\mean{p_x(t)y(t^\prime)}=\frac{\hbar}{4}\left(e^{-i\Omega_+(t-t^\prime)}-e^{-i\Omega_-(t-t^\prime)}\right),
\end{equation}
that disappear in the continous limit $\theta\rightarrow 0$.

\section{4. Perturbative expansion}

The usual perturbative methods to study the typical anharmonic correction in commutative configuration space, i.e. $\lambda x^4$, to order $O(\lambda)$  is based on the evaluation of  (see Eq.~(\ref{pert.exp}))
\begin{equation}\begin{split}
&Z[J_x]=\\&\quad\tilde{Z}_0(t_0,t_*)e^{\left(-i\frac{\lambda}{4} \int_{t_0}^{t_*} d\tau \left(\frac{\hbar\delta}{i\delta J_x(\tau)}\right)^4\right)} e^{-\frac{1
}{2\hbar^2}J_xi\Delta J_x}\\&\simeq\tilde{Z}_0(t_0,t_*)\Bigg[1-i\frac{\lambda}{4} \int_{t_0}^{t_*} d\tau \left(\frac{\hbar\delta}{i\delta J_x(\tau)}\right)^4\\&\hspace{1 cm}+\mathcal{O}(\lambda^2)\Bigg] e^{-\frac{1}{2\hbar^2}J_xi\Delta J_x}\Bigg|_{J_x=0}.\label{pert}
\end{split}\end{equation}
In particular the $\mathcal{O}(\lambda)$ term turns out to be 
\begin{equation}
-3i\frac{\lambda}{4}\int^{t_*}_{t_0} 
[i\Delta(\tau;\tau)]^2 d\tau.
\end{equation}

On the other hand, one could be interested in non-diagonal interaction potentials and the PSP is precisely a method to set up a perturbative expansion for such potentials by introducing the sources
$J_i(t)$. In these cases the general fourth order functional derivative of the partition function associated with $H_0$ is given by
\begin{equation}\begin{split}
&\frac{\hbar^4\delta^4 Z_0[J_i]}{\delta J_a(t)\delta J_b(t^\prime) \delta J_c(t^{\prime\prime})\delta J_d(t^{\prime\prime\prime})}\Bigg|_{J_i=0}=\\&i\Delta_{bd}(t^\prime;t^{\prime\prime\prime})i\Delta_{ac}(t;t^{\prime\prime})+i\Delta_{ad}(t;t^{\prime\prime\prime})i\Delta_{bc}(t^\prime;t^{\prime\prime})\\&\hspace{3.5 cm}+i\Delta_{cd}(t^{\prime\prime};t^{\prime\prime\prime})i\Delta_{ab}(t;t^{\prime}).\end{split}
\end{equation}

As a simple example to outline the perturbative approach in phase space, let us consider
the interaction term for the non-commutative anharmonic oscillator in 2D 
\begin{equation}
H_{\mathrm{int}}=\frac{\lambda}{4} (x^{\prime 2}+y^{\prime 2})^2,
\end{equation}
where for simplicity we suppose
\begin{equation}
\lambda_x=\lambda_y=\frac{\lambda_{xy}}{2}\equiv\lambda
\end{equation}
in eq.(\ref{H int}).
By the substitutions in Eq.~(\ref{cj}), the interaction part $H_{\mathrm{Int}}$ becomes a 4-th order polynomial function in $x,y,p_x$ and $p_y$. 
Following the same procedure to get Eq.~(\ref{pert}), the $\mathcal{O}(\lambda)$ correction can be written as
\begin{equation}\begin{split}\label{52}
&-i\lambda\sum_{n=0}^{4}\int_{t_0}^{t_*}d\tau \left(\frac{\theta}{2\hbar}\right)^n\times\\&f_n\left(\frac{\delta}{\delta J_i(\tau)},\frac{\delta}{\delta J_j(\tau)},\frac{\delta}{\delta J_k(\tau)},\frac{\delta}{\delta J_l(\tau)}\right),
\end{split}\end{equation}
where $f_n$ are 4-th order homogeneous polynomials in the functional derivatives and all odd $n$ terms give no contribution due to the odd powers of the variables in the functional integral. Substituting the explicit expressions for $f_n$ in (\ref{52}), the $\mathcal{O}(\lambda)$ correction is given by
\begin{equation}\begin{split}
&Z[0]=\tilde{Z}[t_0,t_*]\Bigg[1-i\frac{\lambda}{4}
\Big(2\frac{\hbar^2}{(M\Omega)^2}+\theta^2
\\
&\hskip 1cm
+(M\Omega)^2\frac{\theta^4}{8\hbar^2}\Big)
  (t_*\!-\!t_0)
+\mathcal{O}(\lambda^2)\Bigg]
,
\end{split}\end{equation}
consistent with the result obtained by standard perturbative techniques~\cite{Muthukumar:2002}.

Another interesting case is that of a potential dependent on $p^\prime_x$ and $p^\prime_y$ when the phase space geometry is non-commutative. The interaction term will be 
\begin{equation}
H_{\mathrm{int}}=\lambda_x {x'}^4+\lambda_y {y'}^4+\lambda_{p_x}p_x'^4+\lambda_{p_y}p_y'^4.\label{H int'}
\end{equation}
If we substitute eq.(\ref{general omeomorphism}) in eq.(\ref{H int'}) we obtain a 4th order homogenous polynomial expression in the canonical variables $\{x,y,p_x,p_y\}$,
 whose free hamiltonian is~(\ref{H0 shifted theta eta}). 
The next step is to apply the perturbative scheme discussed at the beginning of this section for the interaction term(\ref{H int'}). 
Between the previous case and this one there is no formal difference apart from the variation in the measure of the functional integral (\ref{eq:13})  caused by the substitution~(\ref{general omeomorphism}), whose determinant is,
\begin{equation}
 \left(\frac{1-\frac{\bar\theta\bar\eta}{4\hbar^2}}
       {1+\frac{\bar\theta\bar\eta}{4\hbar^2}}\right)^2
,
\end{equation}
whose inverse square root contributes to the partition function $Z$.

In this case the $\mathcal{O}(\lambda)$ contribution is
\begin{equation}\begin{split}
&\frac{Z[0]}{\tilde Z[t_0,t_*]}=1-\frac{3i}{4}
\frac{(t_*\!-\!t_0)}{1-\left(\frac{\bar\theta\bar\eta}{4\hbar^2}\right)^2}
\\
&
\Bigg\{\frac34\left(\frac{\hbar}{\tilde M \tilde\Omega}\right)^2\!
\Big[\lambda_x\!+\!\lambda_y\!+\!(\lambda_{p_x}\!+\!\lambda_{p_y})\left(\frac{\bar\eta}{2\hbar}\right)^4\Big]
\\
&+\frac{3}{8}\bar\theta^2(\lambda_x\!+\!\lambda_y)\!+\!\frac{3}{8}\bar \eta^2(\lambda_{p_x}\!+\!\lambda_{p_y})
\\
&
+\frac34\left(\hbar\tilde M \tilde\Omega\right)^2\!
\Big[\lambda_{p_x}\!+\!\lambda_{p_y}\!+\!(\lambda_{x}\!+\!\lambda_{y})\left(\frac{\bar\theta}{2\hbar}\right)^4\Big]
\Bigg\}
\\
&
+\mathcal{O}(\lambda_i^2)
.
\end{split}
\end{equation}
\section{5. Conclusions and Outlook}
One of the central results of this work has to be the realisation that strong enough non-commutative parameters can induce condensation of coordinates that resembles {\it spontaneous symmetry breaking}, leading to a constant fictitious magnetic field in the $z$-direction \begin{equation}B_z=-\frac{c}{e}\frac{M^2\Omega^2\theta}{2\hbar}.\end{equation}

Moreover, in this paper a new phase space path integral method  has been formulated and investigated for interacting theories. The proposed perturbative expansion has the advantage of treating a very broad class of interacting Hamiltonians by a straightforward generalization of the usual techniques in configuration space.

The PSP approach has been discussed for an exactly solvable system  and  the perturbative expansion has been explicitly applied to the non-commutative anharmonic oscillator which, in terms of commutative dynamical variables, produces an interaction potential which is a 4-th order polynomial in $x,y,p_x$ and $p_y$. 

Of course, the perturbative approach in phase space can be applied to physical systems with general  momentum and position dependent interactions.

Moreover, the technique studied in this paper is the starting point for developing similar perturbative expansions in quantum field theories. 
The generalization requires subtle modifications but, in principle could be a powerful tool to study perturbative expansion on non-uniform backgrounds, typical of higher order derivative theories
\cite{Gubser:2000cd, Castorina:2003zv}.

\subsection{Aknowledgement}
A.G. thanks Utrecht University for hospitality.
T.P. acknowledges supprot from the D-ITP consortium, 
a program of the NWO that is funded by the Dutch Ministry of 
Education, Culture and Science (OCW). This work is part of the research programme of the Foundation for Fundamental Research on Matter (FOM), which is part of the Netherlands Organisation for Scientific Research (NWO).

\section{Appendix A: Non standard time ordered operators}
\label{Non standard time ordered operators}

We are interested in determining the equation for the time evolution of $\mean{\mathrm{T}[p(t)p(t^\prime)]}.$
Let us define the operator 
\begin{equation}\begin{split}
O_F(t)\equiv \mathrm{T}[p(t)p(t^\prime)]=&\theta(t-t^\prime)p(t)p(t^\prime) +\\&+ \theta(t^\prime-t)p(t^\prime)p(t),\end{split} \end{equation}
and take the derivatite w.r.t. $t$ to obtain 
\begin{equation}\begin{split}
\frac{d}{dt}& O_F(t)= \delta(t-t^\prime)\big[p(t),p(t^\prime)\big]_{t=t^\prime}+\\+&\theta(t-t^\prime)\frac{dp(t)}{dt}p(t^\prime) + \theta(t^\prime-t)p(t^\prime)\frac{dp(t)}{dt}.\end{split}
\end{equation}
The commutator of course vanishes and the second derivative gives
\begin{equation}\begin{split}
&\frac{d^2}{dt^2}O_F(t)=\delta(t-t^\prime)\big[\frac{dp(t)}{dt},p(t^\prime)\big]_{t=t^\prime} +\\&+\theta(t-t^\prime)\frac{d^2p(t)}{dt^2}p(t^\prime) + \theta(t^\prime-t)p(t^\prime)\frac{d^2p(t)}{dt^2}.\end{split}
\end{equation}
By using Hamilton's equations we finally have
\begin{equation}
\left(-\frac{d^2}{dt^2}-\omega^2\right)O_F(t)=i\hbar m\omega^2 \delta(t-t^\prime).
\end{equation}

\section{Appendix B: Evaluation of the transition amplitude}
\label{Evaluation of the transition amplitude}

In this appendix we find a suitable expression for Eq.~(\ref{Zpartition}) by a discretization procedure. For the sake of simplicity we shall suppress the source terms that will be rewritten at the end of the appendix.
Starting with $2$ sub-intervals in $[t_0,t_*],$
\begin{equation}\begin{split}
\braket{p_*,t_*}{q_0,t_0}=&\int dq_1 dq_2\braket{p_*}{q_2}\times\\&\hspace{- 2 cm}\braket{q_2}{U(t_*,t_1)|q_1}\braket{q_1}{U(t_1,t_0)|q_0},\label{58}
\end{split}\end{equation}
where \begin{equation}
U(t,t^\prime)=\mathrm{T}\Big[\exp \Big (-\frac{i}{\hbar} \int_t^{t^{\prime}}H(\tau)d\tau\Big)\Big ].
\end{equation} 
In Eq.~(\ref{58}) we have used two times the identity \[\1=\int dq \ketbra{q}{q}\] and $t_1$ is the central point of the time interval.

Recalling that \[\braket{q}{p}=\frac{e^{ipq/\hbar}}{\sqrt{2\pi\hbar}}\]
we get 
\begin{equation}\begin{split}
\braket{p_*,t_*}{q_0,t_0}&=\int dq_1 dq_2\frac{e^{-ip_*q_2/\hbar}}{\sqrt{2\pi\hbar}}\times\\&\hspace{-1.5 cm}\braket{q_2}{{U}(t_*,t_1)|q_1}\braket{q_1}{{U}(t_1,t_0)|q_0}.\end{split}
\end{equation}
Let's now consider the matrix element \begin{equation}\begin{split}
&\braket{q_2}{{U}(t_2,t_1)|q_1}=\braket{q_2}{e^{-\frac{i}{\hbar}\delta t \frac{{p}^2}{2m}}e^{-\frac{i}{\hbar}\delta t V({q})}|q_1}\\
&=\int dp_1\braket{q_2}{e^{-\frac{i}{\hbar}\delta t \frac{{p}^2}{2m}}|p_1}\braket{p_1}{e^{-\frac{i}{\hbar}\delta t V({q})}|q_1}\\
&\hspace{-0.1 cm}=\!\!\!\int\hspace{-0.2 cm} \frac{dp_1}{2\pi\hbar}\exp\left(-\frac{i}{\hbar}\delta t H(p_1,q_1) + \frac{i}{\hbar}p_1(q_2-q_1)\right),
\end{split}
\end{equation}
where $\delta t$ is $(t_2-t_1),$ or, in general $(t_l-t_{l-1}).$
Therefore for just two intervals, we have 
\begin{equation}\begin{split}
&\braket{p_*,t_*}{q_0,t_0}=\int dq_1dq_2\frac{e^{-i\frac{p_*q_2}{\hbar}}}{\sqrt{2\pi\hbar}}\int \frac{dp_0dp_1}{(2\pi\hbar)^2}\times\\&\exp\Big(\frac{i}{\hbar}\delta t\Big(p_1\frac{(q_2-q_1)}{\delta t}- H(p_1,q_1)+\\&+p_0\frac{(q_1-q_0)}{\delta t}- H(p_0,q_0) \Big)\Big),
\end{split}\end{equation}
and if we consider $N$ intervals we get
\begin{equation}\begin{split}
&\braket{p_*,t_*}{q_0,t_0}=\prod_{l=1}^N\int dq_l\frac{e^{-i\frac{p_*q_N}{\hbar}}}{\sqrt{2\pi\hbar}}\times\\&\braket{q_l,t_l}{q_{l-1},t_{l-1}},
\end{split}\end{equation}
where
\begin{equation}\begin{split}
&\braket{q_l,t_l}{q_{l-1},t_{l-1}}=\int \frac{dp_{l-1}}{\sqrt{2\pi\hbar}}\times\\&\exp\left( \frac{i}{\hbar}\delta t\left(p_{l-1}\frac{q_l-q_{l-1}}{\delta t}-H(q_{l-1},p_{l-1})\right)\right).
\end{split}\end{equation}
After substituting in Eq.~(\ref{58}) one has
\begin{equation}\begin{split}
&\braket{p_*,t_*}{q_0,t_0}=\\&\prod_{l=1}^N\int dq_l\frac{e^{-i\frac{p_*q_N}{\hbar}}}{\sqrt{2\pi\hbar}}\prod_{l=0}^{N-1}\int \frac{dp_{l}}{\sqrt{2\pi\hbar}}\times\\&\exp\left( \frac{i}{\hbar}\delta t\sum_{l=0}^{N-1}\left(p_{l}\frac{q_{l+1}-q_{l}}{\delta t}-H(q_l,p_l)\right)\right)\\
=&\int dq_N\frac{-i\frac{p_*q_N}{\hbar}}{\sqrt{2\pi\hbar}}\prod_{l=1}^{N-1}\int dq_l\prod_{l=0}^{N-1}\int \frac{dp_{l}}{\sqrt{2\pi\hbar}}\times\\&\exp\left( \frac{i}{\hbar}\delta t\sum_{l=0}^{N-1}\left(p_{l}\frac{q_{l+1}-q_{l}}{\delta t}-H(q_l,p_l)\right)\right).
\end{split}\end{equation}
In the limit $N\rightarrow\infty$ one obtains
\begin{equation}\begin{split}\label{b8}
&F_J(p_*,t_*|q_0,t_0)=\int dq_N \frac{e^{-i\frac{p_*q_N}{\hbar}}}{\sqrt{2\pi\hbar}}\times\\&\int_{q(t_0)=q_0}^{q(t_*)=q_N}\mathcal{D}q\mathcal{D}p\exp\Bigg(\frac{i}{\hbar}\int_{t_0}^{t_*}dt\Big(p\dot{q}-H(q,p)+\\&\hspace{4 cm}+J_qq(t)+J_pp(t)\Big)\Bigg),
\end{split}\end{equation}
where the source terms have been again included.

\section*{Appendix C: Solution of the inversion problem}
\label{Solution of the inversion problem}

In this appendix we address the solution of Eq.~(\ref{eq:14}) for $i\Delta_{ij}$ assuming $D^{ij}$ from Eq.~(\ref{eq:12}). For the sake of simplicity we usually omit the time dependence on the functions.
The system defined by the matrix (\ref{eq:14}),
which we can write as
 \begin{equation}\label{eq:19}\Delta_{ij}= \left (
 \begin{array}{cc}
 \Delta_1 & \Delta_2\\
\Delta_3 & \Delta_4 \\
\end{array}
\right ),
\end{equation}
has two types of equations with homogeneous and Dirac-$\delta$ sources. The equations involve either indices 1 and 3 or 2 and 4.
Let's write the equations for the case(1,3):
\begin{equation}\label{eq:20}
\begin{cases}
-m\omega^2\Delta_1(t;t^\prime)-\partial_t\Delta_3(t;t^\prime)=\hbar\delta(t-t^\prime)\\
\partial_t\Delta_1(t;t^\prime)-\frac{1}{m}\Delta_3(t;t^\prime)=0\\
\mathrm{b.c.}\hspace{1 ex}\mathrm{on} \hspace{1 ex} \Delta_1(t;t^\prime).
\end{cases}\end{equation}
From the second equation one has
\begin{equation}\label{eq:21}\partial_t \Delta_1(t;t^\prime)=\frac{1}{m}\Delta_3(t;t^\prime),\end{equation} and by taking the derivative w.r.t. $t$ and substituting $\partial_t \Delta_3(t;t^\prime)$ into the system in Eq.~(\ref{eq:20}) we get 
\begin{equation}\label{eq:22}
\begin{cases}
m(-\partial_t^2-\omega^2)\Delta_1=\hbar\delta(t-t^\prime)\\
\Delta_3(t;t^\prime)=m\partial_t\Delta_1(t;t^\prime)\\
\mathrm{b.c.}\hspace{1 ex}\mathrm{on} \hspace{1 ex} \Delta_1(t;t^\prime).
\end{cases}\end{equation}
From Eq.~(\ref{1}) and the previous expression we know that $\Delta_1$ is a Feynman type propagator: \begin{equation}\label{eq:23}i\Delta_1(t;t^\prime)=\theta (t-t^\prime)i\Delta^+_1(t;t^\prime)+\theta (t^\prime-t)i\Delta^-_1(t;t^\prime)\end{equation}
where $\Delta^\pm_1$ have to satisfy the homogenous equation \begin{equation}\label{eq:24} m(-\partial_t^2-\omega^2)i\Delta^\pm_1(t;t^\prime)=0,\end{equation}
i.e. 
\begin{equation}\label{eq:25} i\Delta_1^\pm(t;t^\prime)=A_1^\pm(t^\prime)e^{-i\omega t}+B_1^\pm(t^\prime)e^{i\omega t}.\end{equation}
One gets analogous results for the system (2,4), which can be reduced to the system of eqs.
\begin{equation}\label{eq:26}
\begin{cases}
(-\partial_t^2-\omega^2)\Delta_4=m\omega^2\hbar\delta(t-t^\prime)\\
\partial_t\Delta_2(t;t^\prime)=-\frac{1}{m\omega^2}\Delta_4(t;t^\prime)\\
\mathrm{b.c.}\hspace{1 ex}\mathrm{on} \hspace{1 ex} \Delta_2(t;t^\prime).
\end{cases}\end{equation}
Again $\Delta_4$ is a Feynman type propagator.

We have now to impose the boundary conditions.
For $\Delta_1(t;t^\prime)$ the boundary conditions are \begin{equation}\label{eq:27}\begin{cases} m\delta(t-t^0)(i\omega-1)\Delta_1(t;t^\prime)=0\\ 
 m\delta(t-t^*)(i\omega+1)\Delta_1(t;t^\prime)=0.
\end{cases}\end{equation}
By calling $\Delta t= t-t^\prime,$ for $t=t^0$ we have for the step function $\theta(\Delta t)=0$ since $t^\prime>t_0$ while in the case $t=t^*,$ $\theta(-\Delta t)=0$ and we get the conditions for $\Delta^+_1$ and $\Delta^-_1,$ namely 
\begin{equation} 
\label{eq:28}\begin{split}
&\delta(t-t^0)(i\omega-1)\Delta^-_1(t;t^\prime)=0,\\& \delta(t-t^*)(i\omega+1)\Delta^+_1(t;t^\prime)=0.
\end{split}\end{equation}
Following the previous analysis in the configuration space in Ref.~\cite{Glavan:2018}, we interpret $i\omega$ as a $\partial_t$ applied to $\Delta_1^-$ and as  $-\partial_t$ if applied to $\Delta^+_1$ and we obtain
\begin{equation}\label{eq:29}\begin{split}
&i\Delta^+_1(t;t^\prime)=A^+_1(t^\prime)e^{-i\omega t},\\& i\Delta^-_1(t;t^\prime)=B^-_1(t^\prime)e^{i\omega t}.
\end{split}
\end{equation}
Finally by deriving Eq.~(\ref{eq:23}) w.r.t. $t$ one gets \begin{equation}\begin{split}
&i\partial_t\Delta_1(t;t^\prime)=\delta(\Delta t) [i\Delta^+_1-i\Delta^-_1]+\\&\theta(\Delta t)i\partial_t\Delta^+_1+\theta(-\Delta t)i\partial_t\Delta^-_1\label{eq:ob}.
\end{split}\end{equation}
Eq.~(\ref{1}) and Eq.~(\ref{eq:ob}) implies
\begin{equation}
\delta(t-t^\prime)[i\Delta^+_1-i\Delta^-_1]=0,
\end{equation}
and a second derivative gives
\begin{equation}\begin{split}\label{ob2}
&\partial_t^2 i\Delta_1 (t;t^\prime)=\\&\delta (t-t^\prime)[i\partial_t\Delta_1^+(t;t^\prime)-i\partial_t\Delta_1^-(t;t^\prime)] +\\&+\theta (t-t^\prime)i\partial_t^2\Delta_1^+(t;t^\prime)+\theta (t^\prime-t)i\partial_t^2\Delta_1^-(t;t^\prime).
\end{split}\end{equation}
By adding and subtracting in the RHS of Eq.~(\ref{ob2}) the term $\omega^2i\Delta_1,$  and using Eq.~(\ref{eq:24}) 
one can handle the terms not proportional to the $\delta(t-t^\prime).$ After substituting in Eq.~(\ref{eq:29}) a direct comparison with Eq.~(\ref{eq:22}) gives
\begin{equation}\label{eq:30}\begin{cases}i\Delta^+_1(t;t^\prime)=\frac{\hbar}{2m\omega}e^{-i\omega(t-t^\prime)}\\
i\Delta^-_1(t;t^\prime)=\frac{\hbar}{2m\omega}e^{i\omega(t-t^\prime)}.\end{cases}\end{equation}

Finally it is possible to obtain $\Delta_3$ by using Eq.~(\ref{eq:21}):
\begin{equation}\label{eq:d3}\begin{split}
&i\Delta_3(t;t^\prime)=\\&-\theta(\Delta t) \frac{i\hbar}{2}e^{-i\omega(t-t^\prime)}+\theta(-\Delta t)\frac{i\hbar}{2}e^{i\omega(t-t^\prime)}.
\end{split}\end{equation}
Analogously we can find $\Delta_4:$ 
\begin{equation}\begin{split}
i\Delta_4(t;t^\prime)=&\theta (t-t^\prime)\frac{\hbar m\omega}{2}e^{-i\omega(t-t^\prime)}+\\&+\theta(t^\prime-t)\frac{\hbar m\omega}{2}e^{i\omega(t-t^\prime)}.\label{eq:38a}\end{split}
\end{equation}
$\Delta_{2}$ can be evaluated from the system in Eq.~(\ref{eq:26}),
\begin{equation}\begin{split}
i\Delta_2(t,t^\prime)&=\theta(\Delta t) \frac{i\hbar}{2}e^{-i\omega(t-t^\prime)}+\\&-\theta(-\Delta t)\frac{i\hbar}{2}e^{i\omega(t-t^\prime)}.
\end{split}\end{equation}

\section{Appendix D. Dynamical solution}

The quantum solutions of system in Eq.~(\ref{EOM H0}) are
\begin{equation}
\begin{split}
 x(t)=&\frac{ x_0}{2}(\cos\Omega_+t+\cos\Omega_-t)+\\
+&\frac{ y_0}{2}(\sin\Omega_+t-\sin\Omega_-t)+
\\ +&
\frac{ p_{x0}}{2M\Omega}(\sin\Omega_+t+\sin\Omega_-t)+
\\+&
\frac{ p_{y0}}{2M\Omega}(\cos\Omega_-t-\cos\Omega_+t);
\\
 y(t)=&\frac{ y_0}{2}(\cos\lambda_+t+\cos\lambda_-t)+
\\+&
\frac{ x_0}{2}(\sin\Omega_-t-\sin\Omega_+t)
\\ &+\frac{ p_{x0}}{2M\Omega}(\cos\Omega_+t-\cos\Omega_-t)+
\\
+&
\frac{ p_{y0}}{2M\Omega}(\sin\Omega_+t+\sin\Omega_-t);\\
 p_x(t)=&\frac{ p_{x0}}{2}(\cos\Omega_+t+\cos\Omega_-t)+
\\+&
\frac{ p_{y0}}{2}(\sin\Omega_+t-\sin\Omega_-t)+
\\-&
\frac{ x_0M\Omega}{2}(\sin\Omega_-t+\sin\Omega_+t)+
\\ +&
\frac{ y_0M\Omega}{2}(\cos\Omega_+t-\cos\Omega_-t);\\
 p_y(t)=&\frac{ p_{y0}}{2}(\cos\Omega_+t+\cos\Omega_-t)+
\\+&
\frac{ p_{x0}}{2}(\sin\Omega_-t-\sin\Omega_+t)
\\ -&
\frac{ y_0M\Omega}{2}(\sin\Omega_-t+\sin\Omega_+t)+
\\+&
\frac{ x_0M\Omega}{2}(\cos\Omega_-t-\cos\Omega_+t).
\end{split}
\end{equation}


\begin{thebibliography}{99}

\bibitem{Julve:1978xn}
  J.~Julve and M.~Tonin,
  ``Quantum Gravity with Higher Derivative Terms,''
  Nuovo Cim.\ B {\bf 46} (1978) 137.
  doi:10.1007/BF02748637


\bibitem{Eliezer:1989cr}
  D.~A.~Eliezer and R.~P.~Woodard,
  ``The Problem of Nonlocality in String Theory,''
  Nucl.\ Phys.\ B {\bf 325} (1989) 389.
  doi:10.1016/0550-3213(89)90461-6

\bibitem{Hata:1988wn}
  H.~Hata,
  ``Quantization of Nonlocal Field Theory and String Field Theory . 1.,''
  Phys.\ Lett.\ B {\bf 217} (1989) 438.
  doi:10.1016/0370-2693(89)90075-0

\bibitem{Hata:1989ie}
  H.~Hata,
  ``{BRS} Invariance and Unitarity in Closed String Field Theory,''
  Nucl.\ Phys.\ B {\bf 329} (1990) 698.
  doi:10.1016/0550-3213(90)90078-R

\bibitem{Liu:2017puc}
  D.~Langlois, H.~Liu, K.~Noui and E.~Wilson-Ewing,
  ``Effective loop quantum cosmology as a higher-derivative scalar-tensor theory,''
  Class.\ Quant.\ Grav.\  {\bf 34} (2017) no.22,  225004
  doi:10.1088/1361-6382/aa8f2f
  [arXiv:1703.10812 [gr-qc]].


\bibitem{Connes:1997cr}
  A.~Connes, M.~R.~Douglas and A.~S.~Schwarz,
  ``Noncommutative geometry and matrix theory: Compactification on tori,''
  JHEP {\bf 9802} (1998) 003
  doi:10.1088/1126-6708/1998/02/003
  [hep-th/9711162].

\bibitem{Seiberg:1999vs}
  N.~Seiberg and E.~Witten,
  ``String theory and noncommutative geometry,''
  JHEP {\bf 9909} (1999) 032
  doi:10.1088/1126-6708/1999/09/032
  [hep-th/9908142].

\bibitem{Seiberg:2000gc}
  N.~Seiberg, L.~Susskind and N.~Toumbas,
  ``Space-time noncommutativity and causality,''
  JHEP {\bf 0006} (2000) 044
  doi:10.1088/1126-6708/2000/06/044
  [hep-th/0005015].


\bibitem{Schmidt:1994iz}
  H.~J.~Schmidt,
  ``Stability and Hamiltonian formulation of higher derivative theories,''
  Phys.\ Rev.\ D {\bf 49} (1994) 6354
   Erratum: [Phys.\ Rev.\ D {\bf 54} (1996) 7906]
  doi:10.1103/PhysRevD.49.6354, 10.1103/PhysRevD.54.7906
  [gr-qc/9404038].


\bibitem{Deser:1974cz}
  S.~Deser and P.~van Nieuwenhuizen,
  ``One Loop Divergences of Quantized Einstein-Maxwell Fields,''
  Phys.\ Rev.\ D {\bf 10} (1974) 401.
  doi:10.1103/PhysRevD.10.401

\bibitem{Stelle:1977ry}
  K.~S.~Stelle,
  ``Classical Gravity with Higher Derivatives,''
  Gen.\ Rel.\ Grav.\  {\bf 9} (1978) 353.
  doi:10.1007/BF00760427


\bibitem{Simon:1990ic}
  J.~Z.~Simon,
  ``Higher Derivative Lagrangians, Nonlocality, Problems and Solutions,''
  Phys.\ Rev.\ D {\bf 41} (1990) 3720.
  doi:10.1103/PhysRevD.41.3720


\bibitem{Muthukumar:2002}
  B.~Muthukumar and P.~Mitra,
  ``Noncommutative oscillators and the commutative limit,''
  Phys.\ Rev.\ D {\bf 66} (2002) 027701
  doi:10.1103/PhysRevD.66.027701
  [hep-th/0204149].



\bibitem{Kimberly:2003hp}
  D.~Kimberly, J.~Magueijo and J.~Medeiros,
  ``Nonlinear relativity in position space,''
  Phys.\ Rev.\ D {\bf 70} (2004) 084007
  doi:10.1103/PhysRevD.70.084007
  [gr-qc/0303067].

\bibitem{Barnaby:2007ve}
  N.~Barnaby and N.~Kamran,
  ``Dynamics with infinitely many derivatives: The Initial value problem,''
  JHEP {\bf 0802} (2008) 008
  doi:10.1088/1126-6708/2008/02/008
  [arXiv:0709.3968 [hep-th]].


\bibitem{Maldacena:2002vr}
  J.~M.~Maldacena,
  ``Non-Gaussian features of primordial fluctuations in single field inflationary models,''
  JHEP {\bf 0305} (2003) 013
  doi:10.1088/1126-6708/2003/05/013
  [astro-ph/0210603].

\bibitem{Prokopec:2012ug}
  T.~Prokopec and J.~Weenink,
  ``Uniqueness of the gauge invariant action for cosmological perturbations,''
  JCAP {\bf 1212} (2012) 031
  doi:10.1088/1475-7516/2012/12/031
  [arXiv:1209.1701 [gr-qc]].


\bibitem{Glavan:2018}
Drazen Glavan and Tomislav Prokopec, 
  ``A Pedestrian Introduction to Non-equilibrium QFT,''
https://www.staff.science.uu.nl/~proko101 (unpublished)



\bibitem{Ryder:1985wq}
  L.~H.~Ryder,
  ``Quantum Field Theory,''

\bibitem{Chataignier:2018kay}
  L.~Chataignier, T.~Prokopec, M.~G.~Schmidt and B.~Świeżewska,
  ``Systematic analysis of radiative symmetry breaking in models with extended scalar sector,''
  arXiv:1805.09292 [hep-ph].

\bibitem{Chataignier:2018aud}
  L.~Chataignier, T.~Prokopec, M.~G.~Schmidt and B.~Swiezewska,
  ``Single-scale Renormalisation Group Improvement of Multi-scale Effective Potentials,''
  JHEP {\bf 1803} (2018) 014
  doi:10.1007/JHEP03(2018)014
  [arXiv:1801.05258 [hep-ph]].

\bibitem{Carroll:2001ws}
  S.~M.~Carroll, J.~A.~Harvey, V.~A.~Kostelecky, C.~D.~Lane and T.~Okamoto,
  ``Noncommutative field theory and Lorentz violation,''
  Phys.\ Rev.\ Lett.\  {\bf 87} (2001) 141601
  doi:10.1103/PhysRevLett.87.141601
  [hep-th/0105082].

\bibitem{Banerjee:2001zi}
	R.~Banerjee, ``Dissipation and noncommutativity in planar quantum mechanics,''
	Mod.Phys.Lett.\ 	{\bf A17} (2002) 631
	doi:10.1142/S0217732302006977
	[hep-th/0106280]


\bibitem{spin.non.commutativity}
	H.~Falomir, J.~Gamboa, M.~Loewe, F.~M\'endez, and J. C.~Rojas, 
	``Spin noncommutativity and the three-dimensional harmonic oscillator,''
	Phys.\ Rev.\ D {\bf 85} (2012) 25009
	doi:10.1103/PhysRevD.85.025009
	[arxiv 1111.0511]
\bibitem{Alavi:2005yr}	
	 S.~A.~Alavi,
	 ``Lamb shift and stark effect in simultaneous space-space and momentum-momentum noncommutative quantum mechanics and theta-deformed su(2) algebra,''
	Mod.Phys.Lett. {\bf A 22} (2007) 377-383
	doi:10.1142/S0217732307018579
	[hep-th/0501215]

\bibitem{Jackiw:2002wd}
	R.~Jackiw, 
	``Observations on noncommuting coordinates and on fields depending on them,''
	Annales Henri Poincare {\bf 4S2} (2003) S913-S919
	doi:10.1007/s00023-003-0971-5
	[hep-th/0212146]
	

	
	
\bibitem{Castorina:2003zv}
  P.~Castorina and D.~Zappala,
  ``Nonuniform symmetry breaking in noncommutative lambda phi**4 theory,''
  Phys.\ Rev.\ D {\bf 68} (2003) 065008
  doi:10.1103/PhysRevD.68.065008
  [hep-th/0303030].
  
 
\bibitem{Gubser:2000cd}
  S.~S.~Gubser and S.~L.~Sondhi,
  ``Phase structure of noncommutative scalar field theories,''
  Nucl.\ Phys.\ B {\bf 605} (2001) 395
  doi:10.1016/S0550-3213(01)00108-0
  [hep-th/0006119].


	 



\end{thebibliography}


\end{document}